\documentclass[12pt,american]{article}
\usepackage[T1]{fontenc}
\usepackage[latin9]{inputenc}
\usepackage{geometry}
\usepackage{babel}
\usepackage{amsmath}
\usepackage{amssymb}
\usepackage{graphicx}
\usepackage{wasysym}
\usepackage[unicode=true,pdfusetitle,
 bookmarks=true,bookmarksnumbered=false,bookmarksopen=false,
 breaklinks=false,pdfborder={0 0 1},backref=false,colorlinks=false]
 {hyperref}

\makeatletter
\@ifundefined{date}{}{\date{}}
\usepackage{soul}
\usepackage{textgreek}

\makeatother

\begin{document}
\title{Pseudo-RNA with parallel aligned single-strands and periodic base
sequence as a new universality class}
\author{R. Dengler \thanks{ORCID: 0000-0001-6706-8550}}
\maketitle
\begin{abstract}
We investigate a field theory for RNA-like polymers with periodic
base sequence GCGCG..., where only single-strands aligned in the same
direction form double strands. The field theory is derived from a
lattice model that incorporates excluded volume effects, base sequence,
and temperature dependent renaturation/denaturation. 

The artificial directionality leads to a novel universality class,
not related to conventional branched polymers and Lee-Yang field theory.
This universality class is unstable against natural pairing, where
oppositely aligned single-strands form double strands. The denaturation/renaturation
transition is a continuous crossover between two $\varphi_{n=0}^{4}$
critical points and the critical point of the new universality class.
\end{abstract}

\section{Introduction}

RNA molecules do not have complementary strands, but there may be
matching segments within the same molecule. This allows the formation
of hairpins like (formula adapted from \cite{Gennes1968})
\begin{verse}
\[
\begin{array}{ccccccccccc}
. & \rightarrow & \mathrm{G} & \mathrm{C} & \mathrm{A} & \mathrm{C} & \rightarrow & . & . & . & \backslash\;\\
. & . & | & | & | & | &  &  &  &  & \;\:\,\downarrow.\\
. & \leftarrow & \mathrm{C} & \mathrm{G} & \mathrm{U} & \mathrm{G} & \leftarrow & . & . & . & /\;
\end{array}
\]
\end{verse}
Guanin(G) binds to cytosin(C) and adenin(A) binds to uracil(U), but
only for oppositely aligned segments. The internal direction ($\rightarrow$)
is defined as 5' to 3' on the deoxyribose backbone. This type of directionality
restricts the topology of allowed polymer conformations. A single
strand can make a half-turn and form a hairpin, but cannot make a
full turn to form a self-energy (with a segment of aligned single
strands ($\Rightarrow$)).

Logically, the duplication of genetic code by splitting and supplementing
DNA sense and antisense strands could also work for double strands
aligned in the same direction, and one can ask the question whether
nature had a choice. This is a complicated issue, but it is possible
to make some progress if one restricts oneself to polymers with periodic
base sequence like GCGCGC... or AUAUAU... (identical with the antisense
sequence).

Long polymers in a good solvent exhibit a long correlation length,
a hallmark of critical phenomena. Critical phenomena can be described
with field theories, a branch of theoretical physics that had its
golden age in the 1970's and 1980's. However, RNA-like molecules have
not been investigated in detail in this context. An exception is the
non-standard field theory of Lässig and Wiese for the freezing transition
of RNA with random base sequence at low temperature \cite{Wiese2009}.
This model is based on work \cite{Bundschuh2002} by Bundschuh and
allows to calculate critical exponents of the transition in an expansion
around two space dimensions. It contains approximations that still
require rigorous justification.

Recently we introduced a conventional field theory for RNA-like molecules
with a periodic base sequence and natural directionality \cite{Dengler2023}.
The molecules are RNA-like in the sense that there is a pairing energy
and an excluded volume interaction. Additional features like helicity
and stiffness are expected to be irrelevant for long polymers in the
renormalization group sense. The model describes the temperature dependent
denaturation-renaturation transition. The critical dimension is eight,
and there is a close relationship with conventional branched polymers
\cite{Lubensky1979} and Lee-Yang field theory \cite{Parisi1981}.
An additional positive pairing energy for single-strands aligned in
the same direction is irrelevant for this universality class \cite{Dengler2023},
indicating that near the critical point oppositely aligned pairing
dominates. This fixed point remains stable at low temperatures, in
the state of renaturation. The real case is compounded by the non-deterministic
base sequence. An approximation would be a random base sequence, but
this significantly complicates the model. In principle, the model
with random base sequence should also reproduce the freezing transition
at low temperatures.

In this work we examine the corresponding pseudo-RNA field theory,
where only GCGCG$\rightarrow$... (or AUAUA$\rightarrow$...) strands
aligned in the same direction can pair to a strand ($\Rightarrow$).
If strands aligned in opposite directions had a positive binding energy,
the model would become unstable. In this artificial case there are
no hairpin diagrams, leading to significantly different results. The
critical dimension for the denaturation-renaturation transition is
$6$, and the model represents a new universality class, unrelated
to conventional branched polymers and the Lee-Yang critical point.
There appear to be no such polymers in nature, but the model is simple,
unique, and of theoretical interest. We also use the occasion to clarify
technical and physical aspects identically occurring in the case of
natural directionality.

\section{The model}

All calculations are based on the field theory with action

\begin{align}
S & =\int_{xs}\tilde{\varphi}_{s}\left(r_{0}-\nabla^{2}+\partial_{s}\right)\varphi_{s}+\tfrac{1}{2}\int_{xss'}\tilde{\psi}_{ss'}\left[\tau_{0}-e^{-\beta E}\nabla^{2}+we^{-\beta E}\left(\partial_{s}+\partial_{s'}\right)\right]\psi_{ss'}\label{eq:Act_s}\\
 & \qquad-g/\sqrt{K_{d}}\int_{xss'}\left\{ \varphi_{s}\varphi_{s'}\tilde{\psi}_{ss'}+\psi_{ss'}\tilde{\varphi}_{s}\tilde{\varphi}_{s'}\right\} \nonumber \\
 & \qquad+u_{\varphi}\int_{x}\left(\int_{s}\tilde{\varphi}_{s}\varphi_{s}\right)^{2}+u_{\psi}\int_{x}\left(\int_{ss'}\tilde{\psi}_{ss'}\psi_{ss'}\right)^{2}+u_{\varphi\psi}\int_{x}\left(\int_{s}\tilde{\varphi}_{s}\varphi_{s}\right)\left(\int_{ss'}\tilde{\psi}_{ss'}\psi_{ss'}\right),\nonumber 
\end{align}
derived from a lattice model in section (\ref{sec:LatticeModel}).
Action (\ref{eq:Act_s}) has a simple structure but contains unusual
details. The variables $s$ and $s'$ measure lengths along the polymers.
Expressions like $\int_{x,s}...=\int\mathrm{d}^{d}x\int_{-\infty}^{\infty}\mathrm{d}s...$
are abbreviations for integrals over space $x$ and length variables.
The field $\varphi\left(\rightarrow\right)$ represents the single-strand
and has one length argument $s$. Length arguments are written as
suffixes, space arguments are omitted. The field $\psi\left(\Rightarrow\right)$
represents the double-strand, and has two length arguments $s,s'$,
incrementing in the same direction. 

Fields with a tilde are associated (outgoing) fields. The first interaction
with coupling constant $g$ has two incoming $\varphi\left(\rightarrow\right)$
and one outgoing $\tilde{\psi}\left(\Rightarrow\right)$, the second
interaction has one incoming $\psi\left(\Rightarrow\right)$ and two
outgoing $\tilde{\varphi}\left(\rightarrow\right)$. The fields $\psi$
and $\tilde{\psi}$ are symmetric in their length variables. The parameter
$w$ is a dimensionless coupling constant. The usual factor $K_{d}=2^{-d+1}\pi^{-d/2}/\Gamma\left(d/2\right)$
simplifies calculations with dimensional regularization. The Boltzmann
factors $e^{-\beta E}$ contain the temperature dependence, a pairing
energy $E>0$ per $\psi\left(\Rightarrow\right)$ monomer favors the
double strand. Only one of the parameters $r_{0}$ and $\tau_{0}=c_{1}e^{-\beta E}-c_{0}$
is independent. The independent quantity controls the degree of re-/denaturation.
The total constant polymer mass in a correlation function like $\left\langle \tilde{\varphi}_{s}\left(x\right)\varphi_{s'}\left(x'\right)\right\rangle $
is $\ell=s'-s$. The excluded volume interactions with coupling constants
$u$ are irrelevant at the upper critical dimension $6$ and can be
ignored, at least near $d=6$.

\subsection{Polymer twisting}

Action (\ref{eq:Act_s}) resembles action integrals of standard universality
classes of critical dynamics and reaction-diffusion equations, but
with length variables replacing time, and some peculiarities. The
integrals over length variables $s$, $s'$ for the interaction say
that the interaction is the same for any combination of length variables.
This translational invariance suggests to use the Fourier transformed
representation, for which we use the convention $f_{s}=\int_{\omega}e^{-i\omega s}f_{\omega}$,
with $\int_{\omega}...=\int\mathrm{d}\omega/\left(2\pi\right)...,$
and we call the variables conjugated to $s$ ``frequencies''. It
then turns out that there occur no frequency integrals, except at
the very end in Fourier transformations back to length space. Technically
``frequencies'' thus more resemble parameters like the ``temperature''
variables $r_{0}$ and $\tau_{0}$ \footnote{The translational invariance in length variables gets lost for a non-periodic
base sequence.}. In particular, in a scaling analysis the characteristic value of
this parameter is $\omega\sim1/\ell$, where $\ell$ is polymer length.

The integral $\tfrac{1}{2}\int_{ss'}$ in the $\tilde{\psi}\psi$
term of action (\ref{eq:Act_s}) extends over all combinations $s,s'$,
but only $\psi_{ss'}$ with $s<s'$ is independent ($s=s'$ cannot
occur). This is compensated by the factor $\tfrac{1}{2}$. This leads
to the first part of the $\psi$ propagator
\begin{equation}
\left\langle \tilde{\psi}_{\nu,\nu'}\psi_{\omega,\omega'}\right\rangle _{0}=\frac{\delta\left(\nu+\omega\right)\delta\left(\nu'+\omega'\right)+\delta\left(\nu'+\omega\right)\delta\left(\nu+\omega'\right)}{v_{k}-iw\left(\omega+\omega'\right)}\left(2\pi\right)^{d+2}\delta^{d}\left(\boldsymbol{p}+\boldsymbol{k}\right),\label{eq:PsiPropSymm}
\end{equation}
the second part follows by symmetry. In eq.(\ref{eq:PsiPropSymm})
one recognizes direct and twisted propagation. 

Some of the twisted or untwisted diagrams contain closed $\varphi\left(\rightarrow\right)$
loops, an example is diagram $\mathrm{D}_{1}''$ in fig.(\ref{fig:Diagrams}).
Such diagrams do not contribute, length variables cannot increment
consistently in a loop. On a more physical level one can start with
a single-strand polymer, which then can form an unlimited number of
conformations. But to get a loop this strand would have to be cut.
Twisting of $\Rightarrow$ only is taken into account in this topological
sense - for long polymers this should suffice. 

\section{Renormalization group calculation}

\begin{figure}
\centering{}\includegraphics[scale=1.6]{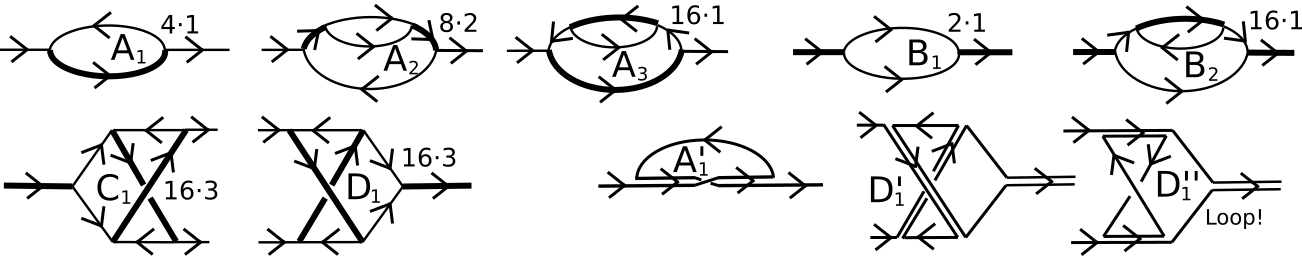}\caption{\label{fig:Diagrams}One- and two-loop contributions to the renormalized
action, with combinatoric factors. The thick line represents double
strand $\psi$. The $\psi$ line could be depicted as in diagrams
$A_{1}'$, $D_{1}'$, $D_{1}''$ and implies untwisted and twisted
propagation with equal weight, eq.(\ref{eq:PsiPropSymm}). The alternatives
lead to the second combinatoric factor.}
\end{figure}
Here, we present the results of a two-loop RG calculation. An essential
part of the formalism has been performed with the computer algebra
framework GiNaC \cite{Ginac2002}. Additional details are provided
in the appendix.

The diagrams are depicted in fig.(\ref{fig:Diagrams}). There is no
one-loop contribution to the $3$-point interactions. This unusual
circumstance excludes any simple equivalence to most other universality
classes. The two-loop flow equations for the renormalized coupling
constants are
\begin{align}
\mu\partial_{\mu}g_{R} & =g_{R}\left[-\tfrac{\epsilon}{2}+g_{R}^{2}\left(2+\tfrac{7}{6}\epsilon\right)-g_{R}^{4}\tfrac{476}{9}\right],\label{eq:FlowEq}\\
\mu\partial_{\mu}w_{R} & =2\left(w_{R}+1\right)\left(w_{R}-\tfrac{1}{2}\right)g_{R}^{2}\left(4+\epsilon\right),\nonumber 
\end{align}
where $\epsilon=6-d$. The calculations have been done in the massless
theory, with a small external wavevector $\mu$. There is an infrared-stable
fixed point
\begin{equation}
g_{*}^{2}=\tfrac{1}{4}\epsilon+\tfrac{217}{144}\epsilon^{2},\qquad w_{*}=\tfrac{1}{2}.\label{eq:FixedPoint}
\end{equation}
The anomalous contributions to the scaling dimensions at the stable
fixed point are 
\begin{align}
\hat{\eta}_{\varphi} & =\mu\partial_{\mu}\ln(Z_{\varphi}/\sqrt{Z_{\omega}})\cong\tfrac{\epsilon}{6}+\epsilon^{2},\label{eq:Dim_Anomal}\\
\hat{\eta}_{\psi} & =\mu\partial_{\mu}\ln(Z_{\psi}/Z_{\omega})\cong\tfrac{\epsilon}{6}+\epsilon^{2},\nonumber \\
\eta_{\omega} & =\mu\partial_{\mu}\ln Z_{\omega}\cong\tfrac{5}{3}\epsilon+\tfrac{199}{18}\epsilon^{2},\nonumber \\
1/\nu_{\varphi}-2 & =-\mu\partial_{\mu}\ln\left(Z_{\varphi^{2}}/Z_{\omega}\right)=\tfrac{2}{3}\epsilon+\tfrac{59}{18}\epsilon^{2}.\nonumber 
\end{align}
The overall scaling dimensions of the $\delta\left(\omega\right)$
factors in eq.(\ref{eq:PsiPropSymm}) are uninteresting, and to get
the usual $k^{2-2\hat{\eta}_{\varphi}}$ and $k^{2-2\hat{\eta}_{\psi}}$
for the two-point vertex functions we have included the factors $Z_{\omega}$
in the definitions of $\hat{\eta}_{\varphi}$ and $\hat{\eta}_{\psi}$
(see appendix). The exponents (\ref{eq:Dim_Anomal}) agree with the
one-loop results \cite{Dengler2020} $\eta_{\varphi}=\hat{\eta}_{\varphi}+\eta_{\omega}/2$,
$\eta_{\psi}=\hat{\eta}_{\psi}+\eta_{\omega}$.

Up to 4-loop order for $w=\tfrac{1}{2}$ and $r_{0}=\tau_{0}=0$ one
finds $\Gamma_{\tilde{\varphi}\varphi}\left(\omega+\omega'\right)=2\Gamma_{\tilde{\psi}\psi}\left(\omega,\omega'\right)$,
which leads to $\hat{\eta}_{\psi}=\hat{\eta}_{\varphi}$ (there are
36 $\Gamma_{\tilde{\varphi}\varphi}$ and 21 $\Gamma_{\tilde{\psi}\psi}$
four-loop diagrams). This equality is unexpected since the number
of $\Gamma_{\tilde{\varphi}\varphi}$ and $\Gamma_{\tilde{\psi}\psi}$
diagrams and their individual multiplicities differ. An example are
diagrams $\mathrm{A}_{2}+\mathrm{A}_{3}$ and $\mathrm{B}_{2}$ of
fig.(\ref{fig:Diagrams}). A generic proof might be recursive in the
number of loops. The analogous equation $\Gamma_{\tilde{\varphi}\varphi}=2\Gamma_{\chi\chi}$
is valid in the case of natural directionality ($\rightleftharpoons$)
\cite{Dengler2023}. Is this equivalence of polymer types at the critical
point a generic property of branched polymers?

The rapid growth of exponent $\eta_{\omega}$ with $\epsilon$ is
unusual, even an extrapolation to $d=5$ makes little sense. The radius
of gyration of a polymer of length $\ell$ for fixed point (\ref{eq:FixedPoint})
would be $\xi\sim\ell^{1/\left(2+\eta_{\omega}\right)}.$ The gaussian
relation is $x\sim\sqrt{\ell}$, which is also the exact result for
double strands ($\rightleftharpoons$) with periodic base sequence
in three dimensions \cite{Parisi1981}. A single- or a double-strand
of given length $\ell$ thus has a much smaller extension in the pseudo-RNA
case (assuming periodic base sequence).

However, the excluded volume interactions, which are strongly irrelevant
at $d=6,$ may become relevant for some $d<6$. Although the model
with pseudo-RNA GCGCG... is hypothetical, it is unique, based on a
lattice model and related to the RNA world, and it would be of interest
to clarify the behavior in three dimensions. 

\section{Temperature dependent crossover}

 Model (\ref{eq:Act_s}) has two thermodynamic degrees of freedom,
total polymer mass $\ell=m_{\varphi}+m_{\psi}$ and degree of denaturation,
i.e. the mass ratio $m_{\varphi}/m_{\psi}$, determined by $r_{0}$
and $\tau_{0}$. Formally $r_{0}$ and $\tau_{0}$ can be considered
as independent. 

A Boltzmann factor $e^{\beta E}$ per $\psi$ monomer is equivalent
to a Boltzmann factor $e^{-\beta E/2}$ per $\varphi$ monomer, which
allows to attribute the temperature dependence to $r_{0}$ or $\tau_{0}$
(See section \ref{subsec:TempDependence}. There occur other factors
$e^{\beta E}$ in action (\ref{eq:Act_s}), but these do not change
substantially near the critical point).

The universality class defined by action (\ref{eq:Act_s}) is realized
on the critical curve $r_{0}=r_{0c}\left(\tau_{0}\right)$ or $\tau_{0}=\tau_{0c}\left(r_{0}\right)$
in the $r_{0}$-$\tau_{0}$ plane. The parameters $r_{0}$ and $\tau_{0}$
always occur in combinations $r_{0}-i\omega$ and $\tau_{0}-i\omega$.
The typical value of $\omega$ is $1/\ell$, which thus is the appropriate
unit to measure $\Delta r_{0}$ and $\Delta\tau_{0}$ near the critical
point. The masses $m_{\varphi}$ and $m_{\psi}$ do not follow a power
law in $\Delta r_{0}$ or $\Delta\tau_{0}$ near the critical point
because $\Delta r_{0}-i\omega$ and $\Delta\tau_{0}-i\omega$ remain
finite. Instead $m_{\varphi}/m_{\psi}$ reaches a finite value.

But away from the critical curve for $\Delta r_{0}>1/\ell$ or $\Delta\tau_{0}>1/\ell$
the masses $m_{\varphi}$ or $m_{\psi}$ diminish according to a power
law in a range of many $1/\ell$. This temperature range becomes small
for long polymers. The exponent can be deduced from a situation with
(originally) a $\varphi$ polymer of length $\ell$, 
\begin{align*}
m_{\varphi} & \propto-\partial_{r_{0}}\ln\left\langle \tilde{\varphi}_{0}\left(0\right)\int_{x}\varphi_{\ell}\left(x\right)\right\rangle \cong\ell/\left(2+c_{\varphi}\ell\Delta r_{0}\right),\\
m_{\psi} & \propto-\partial_{\tau_{0}}\ln\left\langle \tilde{\varphi}_{0}\left(0\right)\int_{x}\varphi_{\ell}\left(x\right)\right\rangle \cong\ell/\left(2+c_{\psi}\ell\Delta\tau_{0}\right).
\end{align*}
The derivative implies the power law in $\ell\Delta r_{0}\gg1$ or
$\ell\Delta\tau_{0}\gg1$, the constant $2$ in the denominator has
been added to interpolate to $m_{\varphi}=m_{\psi}=\ell/2$ on the
critical manifold. Note that $\Delta r_{0}$ grows if temperature
is diminished. 

At low temperature, far away from the critical point (\ref{eq:FixedPoint}),
there remains the pure $\psi$ system equivalent to the $\Phi_{n=0}^{4}$
model. At high temperatures there remains the pure $\varphi$ system
also equivalent to the $\Phi_{n=0}^{4}$ model.

It is of interest to compare this with the case of RNA with base sequence
GCGCG... or AUAUA... and natural directionality. Here the double polymer
$\chi\left(\rightleftharpoons\right)$ interacts with itself via a
$\chi^{3}$ interaction, and in the low-temperature limit there remains
the universality class of the conventional branched polymer \cite{Lubensky1979},
which also describes the intermediate range with $m_{\varphi}\cong m_{\psi}$
\cite{Dengler2023}. An analogous pseudo-RNA $\psi^{3}(\Rightarrow)$
interaction is topologically forbidden.

\section{Derivation of field theory from lattice model}

\label{sec:LatticeModel}A derivation of the $\psi$ sector ($\Rightarrow$)
of field theory (\ref{eq:Act_s}) from a lattice model is straightforward,
of interest are the temperature dependence and a variable base sequence.
The derivation of the $\varphi$ sector is similar, it uses one length
variable instead of two. 

Assume that for every lattice point $i$ of a rectangular lattice
there are operators $b_{i}^{\mu\nu}=b_{i}^{\nu\mu}$ and $\bar{b}_{i}^{\mu\nu}=\bar{b}_{i}^{\nu\mu}$
carrying two length indices, $\mu,\nu\in\mathbb{Z}$. All operators
commute and $b_{i}^{\mu\nu}b_{i}^{\rho\tau}=\bar{b_{i}}^{\mu\nu}\bar{b_{i}}^{\rho\tau}=0$.
There is an ``average'' with $\left\langle b\right\rangle =\left\langle \bar{b}\right\rangle =0$
and $\left\langle \bar{b}_{i}^{\mu\nu}b_{j}^{\rho\tau}\right\rangle =\delta_{ij}\left(\delta_{\mu\rho}\delta_{\nu\rho}+\delta_{\mu\tau}\delta_{\nu\rho}\right)$.
We are interested in the partition sum

\[
Z=\left\langle \prod_{i,j;\mu,\nu}e^{\tfrac{1}{2}\bar{b}_{i}^{\mu+1,\nu+1}v_{i,j}b_{j}^{\mu,\nu}}\right\rangle =\left\langle \prod_{i,j}\left(1+\tfrac{1}{2}\sum_{\mu\nu}\bar{b}_{i}^{\mu+1,\nu+1}v_{ij}b_{j}^{\mu\nu}\right)\right\rangle ,
\]
where $v_{ij}=1$ for next neighbors and $v_{ij}=0$ else. To verify
that $Z$ counts non-overlapping paths collect the factors containing
lattice point $0$,

\begin{align*}
Z & =\left\langle ...\prod_{m,n}\left(1+\tfrac{1}{4}\sum_{\mu\nu\rho\tau}\bar{b}_{0}^{\mu+1,\nu+1}b_{m}^{\mu\nu}\bar{b}_{n}^{\rho+1,\tau+1}b_{0}^{\rho\tau}\right)\right\rangle =\left\langle ...\prod_{m,n}\left(1+\tfrac{1}{2}\sum_{\mu\nu}\bar{b}_{n}^{\mu+2,\nu+2}b_{m}^{\mu\nu}\right)\right\rangle .
\end{align*}
Here $m$ and $n$ denote neighbors of lattice point $0$, and $b_{0}$
and $\bar{b}_{0}$ must occur in pairs. The remaining $Z$ has the
same form as the original one, but lattice point $0$ is missing.
Instead all paths from neighbor $m$ to $n$ are allowed, but only
one of them at a time. The two length variables increment by two in
such a path. The factor $1/2$ is uninteresting. In total, $Z$ counts
all self-avoiding paths of double strands of the $\Rightarrow$ type.
All closed loops contain inconsistent length variables, and to have
any real paths requires external sources and sinks, factors in $Z$
like $\bar{b}_{i}^{\mu\nu}$ and $b_{i}^{\mu\nu}$.

\subsection{Hubbard-Stratonovich transformation}

With a $\mathbb{Z}^{2}\times\mathbb{Z}^{2}$ matrix $V^{\mu\nu;\rho\tau}=\delta_{\mu,\rho+1}\delta_{\nu,\tau+1}$
one can write 

\begin{align}
Z & =\left\langle e^{\tfrac{1}{2}\bar{b}Vvb}\right\rangle =\left\langle \int_{-i\infty}^{i\infty}\mathrm{D}\tilde{\psi}\int_{-\infty}^{\infty}\mathrm{D}\psi e^{-\tfrac{1}{2}\tilde{\psi}V^{-1}v^{-1}\psi+\tfrac{1}{2}\left(\tilde{\psi}b+\bar{b}\psi\right)}\right\rangle \label{eq:Z_lattice}\\
 & =\int\mathrm{D}\tilde{\psi}\mathrm{D}\psi e^{-\tfrac{1}{2}\sum_{ij\mu\nu}\tilde{\psi}_{i}^{\mu\nu}\left(v^{-1}\right)_{ij}\chi_{j}^{\mu+1,\nu+1}+\sum_{i}\ln\left(1+\tfrac{1}{2}\sum_{\mu\nu}\tilde{\psi}_{i}^{\mu\nu}\psi_{i}^{\mu\nu}\right)}\equiv\int\mathrm{D}\tilde{\psi}\mathrm{D}\psi e^{-S_{\psi}},\nonumber 
\end{align}
where in the second line the average $\left\langle \exp\tfrac{1}{2}\left(\tilde{\psi}b+\bar{b}\psi\right)\right\rangle $
has been evaluated. The next neighbor matrix can be written in the
form $v_{ij}=2d\delta_{ij}+a^{2}\nabla_{ij}^{2},$ where $d$ is space
dimension, $a$ the lattice constant and $\nabla_{ij}^{2}$ the discrete
Laplace operator. The expansion of the exponent of eq.(\ref{eq:Z_lattice})
in $a$ (apart from constant factors) then reproduces the $\psi$
part of action (\ref{eq:Act_s}) with excluded volume interaction
$u_{\psi}$, but still without temperature dependence.

The $\psi$-$\varphi$ three-point interaction is an additional detail
which must be added manually at the level of the lattice model or
the action. If $a_{i}^{\mu}$ and $\bar{a}_{i}^{\mu}$ denote the
operators of the $\varphi$ field then at the lattice level one could
add factors like $\left(1+g\bar{b}_{i}^{\mu\nu}a_{i'}^{\mu}a_{i''}^{\nu}\right)$
and $\left(1+g\bar{b}_{i}^{\mu\nu}a_{i'}^{\mu}a_{i''}^{\nu}\right)$
to $Z$, where $i'$ and $i''$ are neighbors of $i.$ A strictly
local interaction cannot be used because of $a_{i}^{\mu}a_{i}^{\nu}=0$.

\subsection{Base sequence and temperature dependence}

\label{subsec:TempDependence}In the most general case there is a
Boltzmann factor $e^{\beta E_{\mu\nu}}$ for each $\psi$ monomer,
where a binding energy $E_{\mu\nu}>0$ favors $\psi$. The dependence
on length variables becomes a dependence on the base sequence $b_{\mu}\in\left\{ 1,2,3,4\right\} $
for a given polymer conformation, i.e. $E_{\mu\nu}=E\left(b_{\mu},b_{\nu}\right)$.
The partition sum then contains $\hat{V}^{\mu\nu;\rho\tau}=\delta_{\mu,\rho+1}\delta_{\nu,\tau+1}e^{\beta E_{\rho\tau}}$
with $\left(\hat{V}^{-1}\right)^{\mu\nu;\rho\tau}=\delta_{\mu,\rho-1}\delta_{\nu,\tau-1}e^{-\beta E_{\mu\nu}}$
instead of $V$, and the action becomes
\begin{align*}
S_{\psi} & =\tfrac{1}{2}\sum_{ij\mu\nu}\tilde{\psi}_{i}^{\mu\nu}v_{ij}^{-1}e^{-\beta E_{\mu\nu}}\psi_{j}^{\mu+1,\nu+1}-\sum_{i}\ln\left(1+\tfrac{1}{2}\sum_{\mu\nu}\tilde{\psi}_{i}^{\mu\nu}\psi_{i}^{\nu\mu}\right)\\
 & \cong\tfrac{1}{2}\sum_{ij\mu\nu}\tilde{\psi}_{i}^{\mu\nu}e^{-\beta E_{\mu\nu}}\left(A-B\nabla_{ij}^{2}+\left(\partial_{\mu}+\partial_{\nu}\right)\right)\psi_{j}^{\mu,\nu}-\sum_{i}\ln\left(1+\tfrac{1}{2}\sum_{\mu\nu}\tilde{\psi}_{i}^{\mu\nu}\psi_{i}^{\mu\nu}\right).
\end{align*}
This shows that in principle the model allows to take an arbitrary
base sequence into account. The case of a periodic base sequence corresponds
to $E_{\mu\nu}=E.$

In simple situations one starts with a single $\varphi$$\left(\rightarrow\right)$
polymer of given length $\ell$ which then forms conformations with
$\varphi$ and $\psi$ segments with constant total mass $\ell$.
Adding in the complete lattice model (with $\varphi$ and $\psi$)
a factor $e^{c}$ for each $\varphi$ monomer and a factor $e^{2c}$
for each $\psi$ monomer (i.e. to the respective $v_{ij}$) only adds
a constant factor to the partition sum. These factors allow to shift
the factor $e^{\beta E}$ from $S_{\psi}$ to $S_{\varphi}$, but
in effect there only are two degrees of freedom, total length and
mass ratio. 

\section{Conclusions}

There are unique field theories for RNA with a simple periodic base
sequence in a good solvent, in the case of natural ($\rightleftharpoons$)
and artificial ($\Rightarrow$) pairing. Renormalization group calculations
provide the following picture:

A large total polymer length $\ell$ implies a large correlation length,
caused by the single-strand state at high temperature or the double-strand
state at low temperature (with no freezing transition for periodic
base sequences).

Moving away from these limiting cases there appears the other polymer
type, first for short wavelengths and then also for longer ones, until
the correlation lengths, anomalous dimensions and masses of the two
polymer types coincide at a critical point. The denaturation - renaturation
transition is continuous, but with the $\psi\varphi^{3}$ critical
point (\ref{eq:FixedPoint}) in between. It occurs in a temperature
range proportional to $1/\ell$, and may resemble a discontinuous
transition for long polymers. In the pseudo-RNA case the gyration
radius is considerably smaller than the usual $\xi\sim\sqrt{\ell}$,
though a quantitative result for $d=3$ is difficult to obtain. The
pseudo-RNA case ($\Rightarrow$) is unstable against natural pairing
($\rightleftharpoons$).

Universality classes offer generic insights into the physics, independent
of quantitative results. It would be of interest to understand what
becomes of these universality classes in the case of a non-deterministic
base sequence.

\bibliographystyle{habbrv}
\bibliography{DoublePolymer_Synopsis}
\bigskip{}

\appendix
\addcontentsline{toc}{section}{Appendix}

\section*{Appendix: Renormalization scheme}

The finite factors $e^{-\beta E}$ in action (\ref{eq:Act_s}) can
be ignored in an RG calculation. The integrals of the diagrams listed
in fig.(\ref{fig:Diagrams}) can been evaluated with $\omega=0$.
Integral $\hat{C}_{1}=\hat{D}_{1}=1/\left(2\epsilon\right)+O\left(\epsilon\right)$
can be taken from \cite{Amit1976}. All other integrals are of the
type

\begin{align*}
\Sigma_{\mu\nu}\left(p\right) & =\tfrac{1}{K_{d}}\int\frac{\mathrm{d}^{d}q}{\left(q^{2}+2qp+p^{2}\right)^{\mu}q^{2\nu}}\overset{\checked}{=}S_{\mu\nu}p^{-2\left(\mu+\nu-d/2\right)},\\
S_{\mu\nu} & =\frac{\Gamma\left(\tfrac{d}{2}\right)\Gamma\left(\mu+\nu-\tfrac{d}{2}\right)}{2\Gamma\left(\mu\right)\Gamma\left(\nu\right)}B\left(\tfrac{d}{2}-\nu,\tfrac{d}{2}-\mu\right),\\
A_{\nu_{1}..\nu_{4}}\left(k\right) & =\tfrac{1}{K_{d}}\int\frac{S_{\nu_{3}\nu_{4}}\mathrm{d}^{d}p}{\left(p+k\right)^{2\nu_{1}}p^{2\left(\nu_{234}-d/2\right)}}=\frac{S_{\nu_{3}\nu_{4}}\Gamma\left(\tfrac{d}{2}\right)\Gamma\left(\nu_{1234}-d\right)}{2\Gamma\left(\nu_{1}\right)\Gamma\left(\nu_{234}-\tfrac{d}{2}\right)}B\left(d-\nu_{234},\tfrac{d}{2}-\nu_{1}\right)k^{-2\left(\nu_{1234}-d\right)}.
\end{align*}
where $\nu_{234}=\nu_{2}+\nu_{3}+\nu_{4}$ etc. One also needs the
derivatives

\begin{align*}
\partial_{i\omega}\hat{A}_{1} & =\partial_{i\omega}\Sigma_{11}=\left(1+2w\right)\Sigma_{21},\\
\partial_{i\omega}\hat{A}_{2} & =\partial_{i\omega}A_{1211}=A_{2211}+2A_{1221}+4wA_{1311},\\
\partial_{i\omega}\hat{A}_{3} & =\partial_{i\omega}A_{1211}=2A_{1311}+A_{1221}+2wA_{2211}+2wA_{1212},\\
\partial_{i\omega}\hat{B}_{1} & =\partial_{i\omega}\Sigma_{11}=2\Sigma_{21},\\
\partial_{i\omega}\hat{B}_{2} & =\partial_{i\omega}A_{1211}=A_{2211}+2A_{1311}+A_{1212}+2wA_{1221}
\end{align*}
and similar derivatives with respect to $r_{0}$. The renormalization
scheme is $\varphi=Z_{\varphi}\varphi_{R}$, $\tilde{\varphi}=Z_{\varphi}\tilde{\varphi}_{R},\psi=Z_{\psi}\psi_{R}$,
$\tilde{\psi}=Z_{\psi}\tilde{\psi}_{R}$, $\left(\varphi\tilde{\varphi}\right)=Z_{\varphi^{2}}\left(\varphi\tilde{\varphi}\right)_{R}$
and $s=Z_{\omega}^{-1}s_{R}.$ A suitable set of normalization conditions
(all evaluated at $\omega=0$) is
\begin{align*}
\partial_{k^{2}}\Gamma_{\varphi\tilde{\varphi},k^{2}=\mu^{2}}^{R} & =1, & \partial_{k^{2}}2\Gamma_{\psi\tilde{\psi},k^{2}=\mu^{2}}^{R} & =1,\\
\partial_{\left(-i\omega\right)}\Gamma_{\varphi\tilde{\varphi},k^{2}=\mu^{2}}^{R} & =1, & \partial_{\left(-i\omega\right)}\Gamma_{\psi\tilde{\psi},k^{2}=\mu^{2},\omega=\nu}^{R} & =w_{R},\\
\partial_{r_{0}}\Gamma_{\varphi\tilde{\varphi},k^{2}=\mu^{2}}^{R} & =1,\\
\Gamma_{\psi\tilde{\varphi}\tilde{\varphi},k=\mu}^{R} & =\Gamma_{\varphi\varphi\tilde{\psi},k=\mathrm{\mu}}^{R}=g^{R}.
\end{align*}
The first nonzero contribution to the vertex function $\partial_{\tau_{0}}\Gamma_{\psi\tilde{\psi},k^{2}=\mu^{2}}^{R}$
comes from the two-loop integral $\partial_{\tau_{0}}\hat{B_{2}}$.
This integral contains a divergent subdiagram and is not primitively
divergent, which prevents a simple renormalization.

\pagebreak
\end{document}